\documentstyle[epsfig]{aipproc}

\begin{document}
\title{Characterisation of Hybrid Pixel Detectors with capacitive charge division}

\author{M. Caccia$^*$\thanks{corresponding author; e-mail:
caccia@fis.unico.it}, S.Borghi$^{\dagger}$, R. Campagnolo$^{\dagger}$,M.
Battaglia$^{\ddagger}$, W. Kucewicz$^{\diamond}$, H.Palka$^{\star}$, A. Zalewska$^{\star}$, 
K.Domanski$^{\circ}$,
J.Marczewski$^{\circ}$, D.Tomaszewski$^{\circ}$}

\address{$^*$Universita' degli Studi dell'Insubria, Dip. di Scienze and INFN, Via
Valleggio 11, Como, Italy\\
$^{\dagger}$Universita' degli Studi di MIlano and INFN, Dip. di Fisica,
Via Celoria 16, Milano, Italy\\
$^{\ddagger}$CERN, CH-1211 Geneva 23, Switzerland\\
$^{\diamond}$Univ. of Mining and Metallurgy, Dept. of Electronics, 
al. Mickiewicza 30, Krakow, Poland\\
$^{\star}$High Energy Physics Lab., Institute of Nuclear Physics,
ul. Kawiory 26a, Krakow, Poland\\
$^{\circ}$Institute of Electron Technology, al. Lotnikov, 32/46, PL02468,
Warszawa, Poland}

\maketitle

\begin{abstract}
In order to fully exploit the physics potential of the future high energy
$e^+e^-$ linear collider, a Vertex Tracker providing high resolution track reconstruction 
is required. Hybrid pixel sensors are
an attractive technology due to their fast read-out capabilities and radiation 
hardness. A novel pixel detector layout with interleaved cells between the readout nodes
has been 
developed to improve the single point resolution. The results of the 
characterisation of the first processed prototypes are reported.
\end{abstract}

\section*{Introduction}
The physics programme at future high energy linear colliders, designed to 
deliver $e^+e^-$ collisions at centre-of mass energies $\sqrt{s}$ = 0.3 - 
3~TeV with luminosities in excess to $10^{34}$~cm$^{-2}$~s$^{-1}$, largely 
relies on the ability to identify in a high track density environment 
the flavour of final state fermions with 
high efficiency and purity~\cite{challenges}.
The target impact parameter resolution may be parametrised as  \mbox{5~$\mu$m $\oplus$ $\frac{15~\mu m}
{p_t~(GeV/c)sin^{1/2}(\theta)}$}
and it requires single point resolution at the $5-7~\mu m$ level and a material budget below 
$0.5\%X_{o}$ for each layer. 
This paper summarises the recent results on the development of a novel hybrid pixel
sensor design, tailored to meet the Linear Collider specifications.

\section*{ Hybrid Pixel Sensor Design}
The main limitation of the existing pixel detectors is the achievable single point 
resolution. A resolution below $10~\mu m$ can be obtained by sampling the diffusion of the 
carriers generated along the particle path and adopting an analog read-out to 
interpolate the signals of neighbouring cells. Since the charge diffusion 
r.m.s. in 300~$\mu$m thick silicon is $\simeq$~8~$\mu$m, its efficient sampling
requires a pixel pitch below 50~$\mu$m. As the most advanced read-out 
electronics have a cell dimension of 50 $\times$ 300~$\mu$m$^2$, this 
is limiting the pixel pitch, hence of the resolution.
Future developments in 
deep sub-micron VLSI electronics may help overcoming this limit. However, it is 
interesting to independently explore a sensor design reducing the current constraints.

The proposed pixel detector design exploits a layout, already successfully
adopted in Silicon microstrip detectors, where only one-out-of-$n$ implants 
is read-out. In such a 
configuration, charge carriers generated underneath one of the interleaved
pixel cells induce a signal on the capacitively coupled read-out pixels, leading to 
a spatial accuracy improvement by a proper signal interpolation. Sampling of the 
charge
carrier distribution is achieved by the fine implant pitch and the analog cell size
has to fit the wider read-out pitch. The maximum number of interleaved 
pixels is constrained  by the charge collection mechanism and the two track separation.

Prototypes of detectors with interleaved pixels have been designed and 
manufactured. Thirty-six test structures have been fit on a 
4$^{''}$ wafer, consisting of detectors with 0 to 3 interleaved pixels defining a 
VLSI cell size of
either 200 $\times$ 200~$\mu$m$^2$ or 300 $\times$ 300~$\mu$m$^2$. Details of the layout, 
the technology and the results of the
basic electrostatic tests may be found elsewhere~\cite{woj}-\cite{ken}.

\section*{Detector modelling}
In a simplified model, the detector may be reduced to a 
capacitive network. Each pixel is a node characterised by the backplane capacitance
($C_{bkpl}$) and the
inter-pixel capacitances ($C_{ip}$), dominated by the couplings to the nearest 
and diagonal neighbours.
The $C_{ip}/C_{bkpl}$ ratio is crucial in the detector design, as it defines the 
signal amplitude reduction (an effective charge loss) at the 
output nodes. 

The capacitances of the produced prototypes were measured and the procedure reported 
in ~\cite{ieee}; the results 
are summarised in Table~1, where a comparison 
with a numerical estimate is also shown. 

\begin{table}[h!]
\begin{center}
\caption{\sl Interpixel ($C_{ip}$) and backplane ($C_{bkpl}$) capacitance values 
for different detector structures. }
\begin{tabular}{l c c c c}
\hline
   & chip~1 & chip~2 & chip~3 & chip~4 \\
\hline \hline
Implant width [$\mu$m]     & 100 &  60 &   50 &   34 \\
Implant pitch [$\mu$m]     & 150 & 100 &   75 &   50 \\ 
Readout pitch [$\mu$m]     & 300 & 200 &  300 &  200 \\
No. of pixels in parallel  &  64 & 128 &  126 & 254 \\ \hline
Measured   $C_{ip}$  $[fF]$ & 599$\pm$5 & 1038$\pm$11 & 958$\pm$5 & 
2098$\pm$30\\
Calculated $C_{ip}$  $[fF]$ & 630$\pm$60 & 880$\pm$67 & 690$\pm$70 & 
980$\pm$40\\
Measured   $C_{bkpl}$  $[fF]$ & 447$\pm$5 & 368$\pm$5 & 218$\pm$5 & 
185$\pm$5\\
Calculated $C_{bkpl}$  $[fF]$ & 470$\pm$70 & 410$\pm$90 & 230$\pm$35 & 
211$\pm$60\\ \hline
Total $C_{ip}$ [$fF$] & 26.2$\pm$4.0 & 13.4$\pm$0.9 & 13.8$\pm$2.0 & 
9.6$\pm$1.0\\
$C_{nearest~neighbour}$  [$fF$] & 4.4$\pm$0.7 & 2.0$\pm$0.1 & 2.1$\pm$0.3 & 
1.5$\pm$0.2\\
$C_{bkpl}$ [$fF$] & 7.3$\pm$1.1 & 3.2$\pm$0.7 & 1.9$\pm$0.7 & 
0.8$\pm$0.2\\ \hline
Max. charge loss & 35\%  & 32\% & 77\%  & 67\% \\ 
\hline
\end{tabular} 
\end{center}
\label{tab:capi}
\end{table}

The capacitances 
were calculated solving the Laplace equation inside a 5x5 
pixel matrix with a finite element analysis performed using the 
OPERA-3D package~\cite{opera}. The comparison 
with the calculated values is fair for all of the structures but 
chip 4, where difficulties in the 
simulation were expected because of the small pitch. In 
particular, the maximum number of elements in OPERA-3D was preventing 
an optimal interpixel mesh and the extension to a larger pixel matrix, crucial for a 
proper interpixel capacitance evaluation for pixels with $50 \mu m$ 
pitch in a $350 \mu m$ thick detector. 
The numerical estimate is 
essential to break down the measurements in the single inter-pixel 
contributions, used to specify the capacitive network.
The calculated single pixel main capacitances are also summarised 
in Table~1, together with the maximum charge loss estimated by a network analysis with a
dedicated software based on the node potential method. The role of $C_{ip}$ may be stressed
referring to chip 4, where the maximum charge loss is reduced to 51\% if the measured
interpixel capacitances are assumed.


\section*{Charge collection studies}
The achievable resolution is proportional to the pixel pitch and the Noise over Signal 
ratio (N/S). If the implant pitch is comparable to the diffusion width, the collected charge
at the output nodes will have both the contributions by the direct diffusion and the linear
share by the capacitive coupling. In such a case, an improvement beyond the binary value 
given by the $implant~pitch/\surd 12$ may be expected. The N/S in pixel detectors
can easily exceeds 1/100, due to the small detector capacitance. Because of this, an
effective signal interpolation may be expected even with  $\simeq 50\%$ signal reduction. 

The charge collection properties and the achievable resolution have been directly studied
by shining an infrared diode spot on the backplane of a structure with 
60~$\mu$m implant width, 100~$\mu$m implant pitch and 200~$\mu$m read-out 
pitch. At the diode wavelength of $\lambda$ = 880~nm, the penetration depth in the 
silicon substrate corresponds to 10~$\mu$m. The IR light has been focused to a
spot size of $\simeq$~80~$\mu$m and its position in the detector 
plane controlled by a 2-D stage allowing to scan the pixel array with 
micro-metric accuracy. 
A sketch of the tested structure and the scan direction is displayed in Fig.~1

\begin{figure}[hb!]
\begin{center}
\epsfig{file=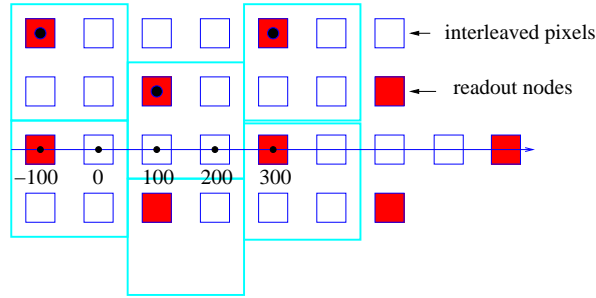,bbllx=0,bblly=-40,bburx=515,bbury=363,height=5.5cm}
\caption{\sl A sketch of the tested structure, with 
 100~$\mu$m implant pitch and 200~$\mu$m read-out pitch. The horizontal line identifies
 the scan direction and defines the coordinate system. The frames define the footprint of an
 hypothetical VLSI cell.}
\end{center}
\label{fig:scan_0}
\end{figure}

A matrix of 4 $\times$ 7 
read-out pixels has been wire-bonded to a VA-1 chip~\cite{ideas}. For each spot position, 
1000 events have been recorded. 
The common mode, pedestal and noise calculation has been initialised for the 
first 300 events. In the subsequent events light was injected every 10 events,
allowing for continuous pedestal tracking. 
Because of the limited data volume, no on-line suppression has been applied 
and the data reduction and cluster search has been performed off-line. 
Results have been averaged over the 70 recorded light pulses, with a peak pulse height
corresponding to N/S $\simeq$ 1/100 and a maximum charge loss of $\simeq$ 40\%, in agreement
with the network analysis. 
The charge sharing may be characterised by the $\eta$ function, 
defined as \mbox{$\eta = \frac{PH_i}{PH_{cluster}}$},
where $PH_i$ is the pulse height on 
the reference pixel i, normalised to the cluster pulse height.
The $\eta$ function by construction ranges in the [0;1] interval and it has a period
equals to the readout pitch. The measured distribution is shown in Fig.2a, where 
the reference pixel for the first period has a centre at x = 100~$\mu m$.
For the structure under test, the ratio between the spot size and the pitch $\simeq$ 0.8
and the experimental $\eta$ curve 
can be understood as a superposition of the effects due to the 
diffusion of the charge carriers created by the IR spot on the 
neighbouring junctions and by the capacitive charge sharing. 
The $\eta$ parametrisation allows a coordinate reconstruction on an event by event basis and
the measurement of the resolution, obtained comparing the laser spot position by the
micrometric stage to the reconstructed values. The results are shown in Fig.~ 2a and 2b.
In the interleaved pixels, where the charge sharing is most efficient, the resolution 
approaches $\simeq 3~ \mu m$, irrespective of the $\simeq$ 40\% charge loss. On the other hand, 
when the charge sharing is minimal the resolution degrades to $\simeq 10~ \mu m$, even with
a peak pulse height. According to these results, the binary resolution defined by the
$implant~pitch/\surd 12$ can be improved by about a factor 4 for a configuration where the
ratio between the charge carrier cloud r.m.s. and the pixel pitch $\simeq$ 0.8. A similar
scaling factor can be expected for a minimum ionising particle detected by a pixel sensor
with $20-25 \mu m$ pitch, as long as the peak $N/S \le 1/100$ and the charge loss is $\le
50\%$. This would lead to intrinsic resolutions $\simeq 2 \mu m$, well within the
specifications for an experiment at the future linear collider.

\begin{figure}[hb!]
\begin{center}
\epsfig{file=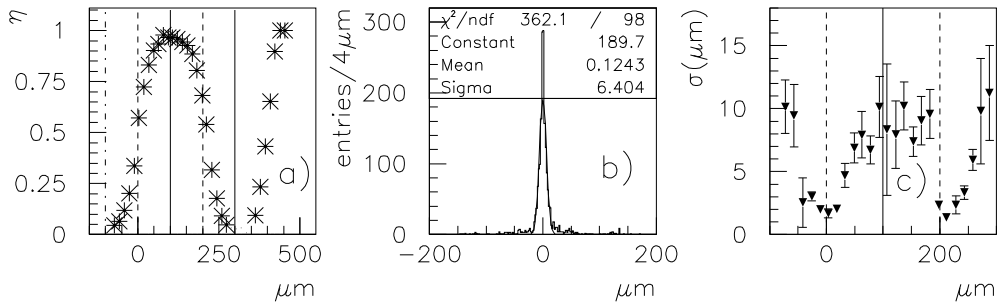,bbllx=13,bblly=345,bburx=519,bbury=533,height=5.0cm}
\caption{\sl The charge sharing among neighbouring read-out pixels, measured by 
$\eta = \frac{PH_i}{PH_{tot}}$ during the detector 
scan (a). The average resolution and the achieved resolution vs. the spot position are 
shown in b)
and c). The vertical lines identify the centre of the pixels; the full line identify the
reference pixel for the $\eta$ calculation}
\end{center}
\label{fig:scan_1}
\end{figure}

\section*{Conclusions}
Prototype pixel detectors with interleaved pixel cells, aimed at improving 
their single point resolution to match the requirements for applications at 
the future linear collider,
have been designed and manufactured. The results of their electrostatic 
characterisation and the preliminary charge collection studies have confirmed
the validity of this detector concept. A second prototype production with $20-25 \mu m$
pixel pitch is planned on a short time scale.

\end{document}